/burl@stx null def /BU.S /burl@stx null def def /BU.SS currentpoint /burl@lly exch def /burl@llx exch def burl@stx null ne burl@endx burl@llx ne BU.FL BU.S if if burl@stx null eq burl@llx dup /burl@stx exch def /burl@endx exch def burl@lly dup /burl@boty exch def /burl@topy exch def if burl@lly burl@boty gt /burl@boty burl@lly def if def /BU.SE currentpoint /burl@ury exch def dup /burl@urx exch def /burl@endx exch def burl@ury burl@topy lt /burl@topy burl@ury def if def /BU.E BU.FL def /BU.FL burl@stx null ne BU.DF if def /BU.DF BU.BB [ /H /I /Border [burl@border] /Color [burl@bordercolor] /Action ¡¡ /Subtype /URI /URI BU.L ¿¿ /Subtype /Link BU.B /ANN pdfmark /burl@stx null def def /BU.BB burl@stx HyperBor- der sub /burl@stx exch def burl@endx HyperBorder add /burl@endx exch def burl@boty HyperBorder add /burl@boty exch def burl@topy HyperBorder sub /burl@topy exch def def /BU.B /Rect[burl@stx burl@boty burl@endx burl@topy] def /eop where begin /@ldeopburl /eop load def /eop SDict begin BU.FL end @ldeopburl def end /eop SDict begin BU.FL end def ifelse

# Process Ordering in a Process Calculus for Spatially-Explicit Ecological Models

Anna Philippou and Mauricio Toro
Department of Computer Science, University of Cyprus
{annap, mtoro}@cs.ucy.ac.cy

*This work was carried out during the tenure by the second author of an ERCIM "Alain Bensoussan" Fellowship Programme. The research leading to these results has received funding from the European Union Seventh Framework Programme (FP7/2007-2013) under grant agreement no 246016.

**An extended abstract of this technical report has appeared in the Proceedings of the Second Symposium on Modelling and Knowledge Management for Sustainable Development (MoKMaSD'13).

**Abstract.** In this paper we extend PALPS, a process calculus proposed for the spatially-explicit individual-based modeling of ecological systems, with the notion of a *policy*. A policy is an entity for specifying orderings between the different activities within a system. It is defined externally to a PALPS model as a partial order which prescribes the precedence order between the activities of the individuals of which the model is comprised. The motivation for introducing policies is twofold: one the one hand, policies can help to reduce the state-space of a model; on the other hand, they are useful for exploring the behavior of an ecosystem under different assumptions on the ordering of events within the system. To take account of policies, we refine the semantics of PALPS via a transition relation which prunes away executions that do not respect the defined policy. Furthermore, we propose a translation of PALPS into the probabilistic model checker PRISM. We illustrate our framework by applying PRISM on PALPS models with policies for conducting simulation and reachability analysis.

## 1 Introduction

Population ecology is a subfield of ecology that deals with the dynamics of species populations and their interactions with the environment. Its main aim is to understand how the population sizes of species change over time and space. It has been of special interest to conservation scientists and practitioners who are interested in predicting how species will respond to specific management schemes and in guiding the selection of reservation sites and reintroduction efforts, e.g. [22, 33].

One of the main streams of today's theoretical ecology is the *individual-based* approach to modeling population dynamics. In this approach, the modeling unit is that of a *discrete individual* and a system is considered as the composition of individuals and their environment. Since individuals usually move from one location to another, it is common in individual-based modeling to represent space explicitly. There are four different frameworks in which *spatially-explicit individual-based models* can be defined [7]. They differ in the way space and time are modeled: each can be treated either discretely or continuously. The four resulting frameworks have been widely studied in Population ecology and they are considered to complement as opposed to compete with each other.

In this paper, we extend our previous work on a process-calculus framework for the spatially-explicit modeling of ecological systems. Our process calculus, PALPS follows the individual-based modeling and, in particular, it falls in the discrete-time, discrete-space class of Berec's taxonomy [7]. PALPS associates processes with information about their location and their species. The habitat is defined as a graph consisting of a set of locations and a neighborhood relation. Movement of located processes is then modeled as the change in the location of a process, with the restriction that the originating and the destination locations are neighboring locations. In addition, located processes may communicate with each other by exchanging messages upon channels. Communication may take place only between processes which reside at the same location. Furthermore, PALPS may model probabilistic events, with the aid of a probabilistic choice operator, and uses a discrete treatment of time. Finally, in PALPS, each location may be associated with a set of attributes capturing relevant information such as the capacity or the quality of the location. These attributes form the basis of a set of expressions that refer to the state of the environment and are employed within models to enable the enunciation of location-dependent behavior.

The extension presented in this paper is related to the issue of *process ordering* inside each time unit. In particular, simulations carried out by ecologists impose an order on the events that may take place within a model. For instance, if we consider mortality and reproduction within a single-species model, three cases exist: concurrent ordering, reproduction preceding mortality and reproduction following mortality. In *concurrent ordering*, individuals may reproduce and die simultaneously. For *reproduction preceding mortality*, the population first reproduces, then all individuals, including new offspring, are exposed to death. For *reproduction following mortality*, individuals are first exposed to death and, subsequently, surviving individuals are able to reproduce. Ordering can have significant implications on the simulation. Thus, alternatives must be carefully studied before conclusions are drawn.

In order to capture process ordering in PALPS, we define the notion of a *policy*, an entity that imposes an order on the various events that may take place within a system. Formally, a policy, $\sigma$, is defined as a partial order on the set of events in the system where, by writing $(\beta) \in \sigma$, we specify that, whenever there is a choice between executing the activities \$ $and \beta$, $\beta$ is chosen. As a result, a policy is defined externally to a process description. This implies that one may investigate the behavior of a system under different event orderings simply by redefining the desired policy without redeveloping the system's description. To capture policies in the semantics of PALPS we extend its transition relation into a *prioritized* transition relation which prunes away all transitions that do not respect the defined policy.

Furthermore, we present a methodology for analyzing models of PALPS with policies via the probabilistic model checker PRISM [1]. To achieve this, we describe a method for translating models of PALPS with policies into the PRISM language and we prove its correctness. We then apply our methodology on simple examples that demonstrate the types of analysis that can be performed on PALPS metapopulation models via the PRISM tool. By contrasting our results with our previous work of [37], we observe that policies achieve a significant reduction in the size of models and may thus enable the analysis of larger systems.

Various formal frameworks have been proposed in the literature for modeling biological and ecological systems. Similarly to ecosystem modeling, these approaches differ in their treatment of time and space and can be considered as supplements as opposed to rivals of each other as each offers a distinct view and different techniques for analyzing systems. One strand is based, like PALPS, on process calculi, and constitute extensions of calculi such as CCS [26], the $\pi$-calculus [27] and CSP [24]. Examples include WSCCS of [43] which follows the discrete-time approach to modeling but does not include the notion of space. As far as continuous time is concerned, there are various proposals including [19, 13, 23, 17] whereas numerous process calculi have been proposed in the literature to model space including [11, 39, 12]. A different approach towards modeling biological and ecological systems is that of *P systems* [38]. P systems were conceived as a class of distributed and parallel computing inspired by the compartmental structure and the functioning of living cells. P-systems have been extended in various directions and they have been applied to a wide range of applications including the field of ecology [34, 35, 8, 14, 9, 31, 15]. Finally, we mention the calculus of looping sequences [6], and its spatial extension [5] and *cellular automata* [21, 16]. Process calculi has been also applied to the modeling of interactive music systems [55, 61, 50, 60, 2, 56, 51, 30, 48, 45, 47, 49, 4, 54, 46, 52, 53, 44] and ecological systems [57, 36, 59, 37, 58].

Regarding the notion of policies employed in PALPS *with policies*, we point out that they are essentially a type of priorities usually referred to in the process-algebra literature as *static priority relations* (see e.g. [18]) and are similar to the priorities defined for P-Systems. In comparison to related works, as far as we know, PALPS *with policies* is the first spatially-explicit formalism for ecological systems that includes the notion of priority and employs this notion to experiment with different process orderings within process descriptions. Furthermore, via the translation to the PRISM language our framework enables to carry out more advanced analysis of ecological models than just simulation, which is the main approach adopted in the related literature. Possible analysis techniques are those supported by the PRISM tool and include model-checking, reachability analysis as well as computing expected behavior [20].

**Structure of the paper.** The structure of the remainder of the paper is as follows. In Section 2 we present the syntax and the semantics of PALPS with policies. We illustrate the expressiveness of the calculus by providing models of systems involving process ordering in Section 2.4. In Section 3 we present a translation of PALPS into the Markov-decision-process component of the PRISM language. We establish the correctness of the translation and we overview the types of analysis that this translation makes possible on PALPS models. We then apply these techniques on simple examples and we explore the potential of the approach in Section 4. Finally, in Section 5, we conclude with a discussion of future work.

## 2 The Process Calculus

In our calculus, *Process Algebra with Locations for Population Systems* (PALPS), we consider a system as a set of individuals operating in space, each belonging to a certain species and inhabiting a location. This location may be associated with attributes which describe characteristics of the location and can be used to define location-dependent behavior of individuals. Furthermore, individuals who reside at the same location may communicate with each other by communicating upon channels, e.g. for preying, or they may migrate to a new location where they may continue their computation. PALPS may model probabilistic events with the aid of a probabilistic operator and uses a discrete treatment of time.

### 2.1 The Syntax

In this section we formalize the syntax of PALPS which is built based on the following basic entities:

- **S**: a set of species ranged over by **s**, **s**′.
- **Loc**: a set of locations ranged over by $\ell$, $\ell'$. The habitat of a system is then implemented via a relation **Nb**, where $(\ell,\ell') \in$ **Nb** exactly when locations $\ell$ and $\ell'$ are neighbors. For convenience, we use **Nb** as a function and write **Nb**$(\ell)$ for the set of all neighbors of $\ell$.
- **Ch**: a set of channels ranged over by lower-case strings.
- $\Psi$: a set of attributes, ranged over by $\psi$, $\psi'$. We write $\psi_\ell$ for the value of attribute $\psi$ at location $\ell$.

Species and locations are characteristics associated with every individual in a PALPS system. The species characteristic is static whereas the location characteristic is dynamic: as computation proceeds, an individual may change its location from $\ell$ to $\ell'$ with the restriction that $(\ell,\ell') \in \mathbf{Nb}$. In turn, attributes are characteristics associated with locations and they may capture information such as the capacity, the temperature or the quality of the location. They form the basis of the set of expressions of the language which is defined below.

**Expressions.** PALPS employs two sets of expressions: *logical expressions*, ranged over by $e$, and *arithmetic expressions*, ranged over by $w$. These expressions are intended to capture environmental situations which may affect the behavior of individuals. Expressions $e$ and $w$ are constructed as follows:

$$e ::= true \mid \neg e \mid e_1 \wedge e_2 \mid w \bowtie c$$
$$w ::= c \mid \psi @\ell^* \mid \mathbf{s}@\ell^* \mid @\ell^* \mid \mathbf{op}_1(w) \mid \mathbf{op}_2(w_1, w_2)$$

where $c$ is a real number, $\bowtie \in \{=, \leq, \geq\}$, $\ell^* \in \mathbf{Loc} \cup \{myloc\}$, and $\mathbf{op}_1$ and $\mathbf{op}_2$ are the usual unary and binary arithmetic operations on real numbers.

To begin with, logical expressions $e$ are built using the propositional calculus connectives, as well as comparisons between an arithmetic expression $w$ and a constant $c$ (e.g., $\mathbf{s}_1@\ell + \mathbf{s}_2@\ell > 1$). Arithmetic expressions include three special expressions interpreted as follows: Expression $\psi@\ell^*$ is equal to the value of attribute $\psi$ at location $\ell^*$. Expression $\mathbf{s}@\ell^*$ is equal to the number of individuals of species $\mathbf{s}$ at location $\ell^*$, and expression $@\ell^*$ denotes the total number of individuals of all species at location $\ell^*$.

Location $\ell^*$ can be an arbitrary location or the special location *myloc*. This latter label is employed to bestow individuals with the ability to express conditions on the status of their current location no matter where that might be, as computation proceeds. Specifically, *myloc* refers to the actual location of the individual in which the expression appears and it is instantiated to this location when the condition needs to be evaluated (see rule (Cond) in Table 3). In conclusion, arithmetic expressions are the set of all expressions formed by arbitrary constants $c$, quantities $\psi@\ell^*$, $\mathbf{s}@\ell^*$, $@\ell^*$, and the usual unary and binary arithmetic operations. Logical expressions and arithmetic expressions are evaluated within a system environment (as defined in Tables 1 and 2).

**Processes.** The syntax of PALPS is given at three levels: (1) the individual level ranged over by $P$, (2) the species level ranged over by $R$, and (3) the system level ranged over by $S$. Their syntax is defined via the following BNFs:

$$P ::= 0 \mid \sum_{i \in I} \eta_i.P_i \mid \bullet\sum_{i \in I} p_i:P_i \mid cond(e_1 \triangleright P_1, \ldots, e_n \triangleright P_n) \mid C$$
$$R ::= !rep.P$$
$$S ::= 0 \mid P:\langle \mathbf{s}, \ell \rangle \mid R:\langle \mathbf{s} \rangle \mid S_1 \mid S_2 \mid S \setminus L$$

where $L \subseteq \mathbf{Ch}$, $I$ is an index set, $p_i \in (0,1]$ with $\sum_{i \in I} p_i = 1$, $e_1, \ldots, e_n$, are logical expressions such that $e_1 \vee \ldots \vee e_n = true$, $C$ ranges over a set of process constants $\mathcal{C}$, each with an associated definition of the form $C \stackrel{def}{=} P$, and

$$\eta ::= a \mid \overline{a} \mid go\,\ell \mid \sqrt{}.$$

Beginning with the individual level, $P$ can be one of the following: Process 0 represents the inactive individual, that is, an individual who has ceased to exist. Process $\sum_{i \in I} \eta_i.P_i$ describes the nondeterministic choice between a set of action-prefixed processes: it can execute any of the activities $\eta_i$ and proceed as the respective $P_i$. We write $\eta_1.P_1 + \eta_2.P_2$ to denote the binary form of this operator. In turn, an activity $\eta$ can be an input action on a channel $a$, written simply as $a$, a complementary output action on a channel $a$, written as $\overline{a}$, a movement action with destination $\ell$, $go\,\ell$, or the time-passing action, written as $\sqrt{}$. Actions of the form $a$, and $\overline{a}$, $a \in \mathbf{Ch}$, are used to model arbitrary activities performed by an individual; for instance, eating, preying and reproduction.

The tick action $\sqrt{}$ measures a tick on a global clock. These time steps are abstract in the sense that they have no defined length and, in practice, $\sqrt{}$ is used to separate the rounds of an individual's behavior.

Process $\bullet\sum_{i \in I} p_i:P_i$ represents the probabilistic choice between processes $P_i$, $i \in I$. The process randomly selects an index $i \in I$ with probability $p_i$, and then evolves to process $P_i$. We write $p_1:P_1 \oplus p_2:P_2$ for the binary form of this operator. The conditional process $cond\,(e_1 \trianglelefteq P_1, \ldots, e_n \trianglelefteq P_n)$ presents the conditional choice between a set of processes: it behaves as $P_i$, where $i$ is the smallest integer for which $e_i$ evaluates to *true*. Note that this choice is deterministic. Finally, process constants provide a mechanism for including recursion in the calculus.

Moving on to the species level, we employ the special *species* process $R$ defined as $!rep.P$. This process is a replicated process which may always receive input through channel *rep* and create new instances of process $P$, where $P$ is a new individual of species $R$. Such inputs will be provided by individuals in the phase of reproduction via the complementary action $\overline{rep}$.

Finally, population systems are built by composing in parallel located individuals and species. An individual is defined as $P:\langle \mathbf{s},\ell \rangle$, where $\mathbf{s}$ and $\ell$ are the species and the location of the individual, respectively. A species is given by $R:\langle \mathbf{s} \rangle$, where $\mathbf{s}$ is the name of the species. Finally, $S\setminus L$ models the restriction of the use of channels in set $L$ within $S$. As a syntactic shorthand, we will write $P:\langle \mathbf{s},\ell,n \rangle$ for the parallel composition of $n$ copies of process $P:\langle s,\ell \rangle$.

## 2.2 The Unprioritized Semantics

The semantics of PALPS is defined in terms of a *structural operational semantics* given at the level of configurations of the form $(E,S)$, where $E$ is an *environment* and $S$ is a population system. The environment $E$ is an entity of the form $E \subset \mathbf{Loc} \times \mathbf{S} \times \mathbb{N}$, where each pair $\ell$ and $\mathbf{s}$ is represented in $E$ at most once and where $(\ell,\mathbf{s},m) \in E$ denotes the existence of $m$ individuals of species $s$ at location $\ell$. The environment $E$ plays a central role in evaluating expressions.

The satisfaction relation for logical expressions $\models$ is defined inductively on the structure of a logical expression, as shown in Table 1. It depends on the evaluation function for arithmetic expressions $val\,(E,w)$ defined in Table 2.

**Table 1: The satisfaction relation for logical expressions**

| | | |
|---|---|---|
| $E \models true$ | always | |
| $E \models \neg e$ | if and only if | $\neg(E \models e)$ |
| $E \models e_1 \wedge e_2$ | if and only if | $E \models e_1 \wedge E \models e_2$ |
| $E \models w \bowtie e$ | if and only if | $val\,(E,w) \bowtie e$ |

**Table 2: The evaluation relation for arithmetic expressions**

| | | |
|---|---|---|
| $val\,(E,c)$ | = | $c$ |
| $val\,(E,\psi@\ell)$ | = | $\psi_\ell$ |
| $val\,(E,\mathbf{s}@\ell)$ | = | $n$, $(\ell,\mathbf{s},n) \in E$ |
| $val\,(E,\mathbf{s}@\ell)$ | = | $0$, $(\ell,\mathbf{s},n) \notin E$ |
| $val\,(E, @\ell)$ | = | $\sum_{\mathbf{s} \in \mathbf{S}} val\,(E,\mathbf{s}@\ell)$ |
| $val\,(E,\mathbf{op}_1(w))$ | = | $\mathbf{op}_1(val\,(E,w))$ |
| $val\,(E,\mathbf{op}_2(w_1,w_2))$ | = | $\mathbf{op}_2(val\,(E,w_1), val\,(E,w_2))$ |

Before we proceed to the semantics we define some additional operations on environments that we will use in the sequel:

**Definition 1.** Consider an environment $E$, a location $\ell$ and a species $\mathbf{s}$.

- $E \oplus (\mathbf{s},\ell)$ increases the count of individuals of species $\mathbf{s}$ at location $\ell$ in environment $E$ by 1:

$$E \oplus (\mathbf{s},\ell) = \begin{cases} E' \cup \{(\ell,\mathbf{s},m+1)\} & \text{if } E = E' \cup \{(\ell,\mathbf{s},m)\} \text{ for some } m \\ E \cup \{(\ell,\mathbf{s},1)\} & \text{otherwise} \end{cases}$$

- $E \ominus (\mathbf{s},\ell)$ decreases the count of individuals of species $\mathbf{s}$ at location $\ell$ in environment $E$ by 1:

$$E \ominus (\mathbf{s},\ell) = \begin{cases} E' \cup \{(\ell,\mathbf{s},m-1)\} & \text{if } E = E' \cup \{(\ell,\mathbf{s},m)\}, m > 1 \\ E' & \text{if } E = E' \cup \{(\ell,\mathbf{s},1)\} \\ \bot & \text{otherwise} \end{cases}$$

We may now define the *unprioritized* semantics of PALPS, presented in Tables 3 and 4. This semantics will then be refined into the *prioritized semantics* which takes into account the notion of policies in Section 2.3. The unprioritized semantics is given in terms of two transition relations: the non-deterministic relation $\longrightarrow_n$ and the probabilistic relation $\longrightarrow_p$. A transition of the form $(E,S) \overset{\mu}{\longrightarrow}_n (E',S')$ means that a configuration $(E,S)$ may execute action $\mu$ and become $(E',S')$. A transition of the form $(E,S) \overset{w}{\longrightarrow}_p (E',S')$ means that a configuration $(E,S)$ may evolve into configuration $(E',S')$ with probability $w$. Whenever the type of the transition is irrelevant to the context, we write $(E,S) \overset{\alpha}{\longrightarrow} (E',S')$ to denote either $(E,S) \overset{\mu}{\longrightarrow}_n (E',S')$ or $(E,S) \overset{w}{\longrightarrow}_p (E',S')$. Action $\mu$ appearing in the non-deterministic relation may have one of the following forms:

- $a_{\ell,\mathbf{s}}$ and $\overline{a}_{\ell,\mathbf{s}}$ denote the execution of actions $a$ and $\overline{a}$ respectively at location $\ell$ by an individual of species $\mathbf{s}$.
- $\tau_{a,\ell,\mathbf{s}}$ denotes an internal action that has taken place on channel $a$, at location $\ell$, and where the output on $a$ was carried out by an individual of species $\mathbf{s}$. This action may arise when two complementary actions take place at the same location $\ell$ or when a move action take place from location $\ell$. Note that this information was not included in the semantics of PALPS as presented, e.g., in [37]. It is, however, necessary in PALPS with policies in order to accommodate the enunciation of policies.

**Rules for individuals.** The rules of Table 3 prescribe the semantics of located individuals in isolation. The first four rules define non-deterministic transitions. The fifth axiom defines a probabilistic transition, and the last two rules refer to both the non-deterministic and the probabilistic case. All rules are concerned with the evolution of the individual in question and the effect of this evolution to the system's environment. A key issue in the enunciation of the rules is to preserve the compatibility of $P$ and $E$ as transitions are executed. We consider each rule separately:

**Table 3: Transition rules for individuals**

$$(E,\ 0{:}(s,\ell)) \overset{\checkmark}{\longrightarrow}_n (E,\ 0{:}(s,\ell))$$

$$
\begin{array}{ll}
(\text{Nil}) & \\
(\text{Tick}) & (E, \sqrt{}.P:\langle s,\ell\rangle) \xrightarrow{\sqrt{}}_n (E^{\text{P},\text{s},\ell}, P:\langle s,\ell\rangle) \\
(\text{Act}) & (E, \eta.P:\langle s,\ell\rangle) \xrightarrow{\eta_{\ell,\text{s}}}_n (E^{\text{P},\text{s},\ell}, P:\langle s,\ell\rangle) \qquad \eta \neq go\ \ell', \sqrt{} \\
(\text{Go}) & (E, go\ \ell'.P:\langle s,\ell\rangle) \xrightarrow{\tau_{go,\ell,\text{s}}}_n (((E^{\text{P},\text{s},\ell,\ell'}, P:\langle s,\ell'\rangle)) \qquad (\ell,\ell') \in \mathbf{Nb} \\
(\text{NSum}) & \dfrac{(E, \eta_i.P_i:\langle s,\ell\rangle) \xrightarrow{\mu}_n (E', P_i:\langle s,\ell'\rangle)}{(E, \sum_{i \in I} \eta_i.P_i:\langle s,\ell\rangle) \xrightarrow{\mu}_n (E', P_i:\langle s,\ell'\rangle)} \\
(\text{PSum}) & (E, \bullet\sum_{i \in I} p_i:P_i:\langle s,\ell\rangle) \xrightarrow{p_i}_p (E^{P_i,\text{s},\ell}, P_i:\langle s,\ell\rangle) \\
(\text{Const}) & \dfrac{(E, P:\langle s,\ell\rangle) \xrightarrow{\alpha} (E', P':\langle s,\ell\rangle)}{(E, C:\langle s,\ell\rangle) \xrightarrow{\alpha} (E', P':\langle s,\ell\rangle)} \quad C \stackrel{def}{=} P \\
(\text{Cond}) & \dfrac{(E, P_i:\langle s,\ell\rangle) \xrightarrow{\alpha} (E', P_i':\langle s,\ell'\rangle), E \models e_i \downarrow \ell, E \not\models e_j \downarrow \ell, j < i}{(E, cond(e_1 \triangleright P_1, \ldots, e_n \triangleright P_n):\langle s,\ell\rangle) \xrightarrow{\alpha} (E', P_i':\langle s,\ell'\rangle)}
\end{array}
$$

where $E^{\text{P},\text{s},\ell} = \begin{cases} E \ominus (\text{s},\ell) & \text{if } P = 0 \\ E & \text{otherwise} \end{cases}$

$E^{\text{P},\text{s},\ell,\ell'} = ((E \ominus (\text{s},\ell)) \oplus (\text{s},\ell'))^{\text{P},\text{s},\ell}$

- Axiom (*Nil*) specifies that the 0 process may execute the time consuming action √. This axiom allows for time-progress in a system with inactive individuals.
- Axiom (*Tick*) specifies that a √-prefixed process will execute the time consuming action √ and then proceed as $P$. The state of the new environment depends on the state of $P$. If $P = 0$ then the individual has terminated its computation and it is removed from $E$ (see the definition of $E^{\text{P},\text{s},\ell}$). If $P \neq 0$ then $E$ remains unchanged.
- Axiom (*Act*) specifies that $\eta.P$ executes action $\eta_{\ell,\text{s}}$ and evolves to $P$. Note that the action is decorated by the location and the species of the individual executing the transition to enable synchronization of the action with complementary actions taking place at the same location (see rule (Par2), Table 4). This axiom excludes the cases of $\eta = go\ \ell$ and $\eta = \sqrt{}$ which are treated in separate axioms.
- According to Axiom (*Go*), an individual may change its location. This gives rise to action $\tau_{go,\ell,\text{s}}$ and has the expected effect on the environment $E$. As we have already mentioned, the label $go$, the location $\ell$ and the species $s$ are recorded to enable the enunciation of policies.
- Rule (*NSum*) describes the behavior of a nondeterministic choice: any of the available summands may be selected and executed.
- Rule (*PSum*) expresses the semantics of probabilistic choice: a process is chosen probabilistically leading to the appropriate continuation. If the resulting state of the individual, namely $P_i$, is equal to 0, then the individual is removed from the environment $E$.
- Rule (*Const*) expresses the semantics of process constants in the expected way.
- Finally, rule (*Cond*) stipulates that a conditional process may perform an action of continuation $P_i$ assuming that $e_i \downarrow \ell$ evaluates to true and all $e_j \downarrow \ell, j < i$ evaluate to false. Note that we write $e \downarrow \ell$ for the expression $e$ with all occurrences of *myloc* substituted by location $\ell$.

**Rules for systems.** We may now move on to Table 4 which defines the semantics of system-level operators. The first two rules define the semantics for the replication operator, the next five rules define the semantics of the parallel composition operator, and the last rule deals with the restriction operator.

According to rules (*R_Tick*) and (*R_Rep*), a species process may idle or it may engage in action $rep_{\ell,\text{s}}$ for any location $\ell$ and create a new individual $P:\langle \text{s},\ell\rangle$.

**Table 4: Transition rules for systems**

$$
\begin{array}{ll}
(\text{R\_Tick}) & \dfrac{R = !rep.P:\langle \text{s}\rangle}{(E, R) \xrightarrow{\sqrt{}}_n (E, R)} \\[1em]
(\text{R\_Rep}) & \dfrac{R = !rep.P:\langle \text{s}\rangle, \ell \in \mathbf{Loc}}{(E, R) \xrightarrow{rep_{\ell,\text{s}}}_n (E \ominus (\text{s},\ell), P:\langle \text{s},\ell\rangle | R)} \\[1em]
(\text{Par1}) & \dfrac{(E, S_1) \xrightarrow{\mu}_n (E', S_1'), (E, S_2) \not\xrightarrow{}_p, \mu \neq \sqrt{}}{(E, S_1|S_2) \xrightarrow{\mu}_n (E', S_1'|S_2)} \\[1em]
(\text{Par2}) & \dfrac{(E, S_1) \xrightarrow{a_{\ell,\text{s}}}_n (E_1, S_1'), (E, S_2) \xrightarrow{\overline{a}_{\ell,\text{s}'}}_n (E_2, S_2')}{(E, S_1|S_2) \xrightarrow{\tau_{a,\ell,\text{s}'}}_n (E \otimes (E_1, E_2), S_1'|S_2')} \\[1em]
(\text{Par3}) & \dfrac{(E, S_1) \xrightarrow{w_1}_p (E_1, S_1'), (E, S_2) \xrightarrow{w_2}_p (E_2, S_2')}{(E, S_1|S_2) \xrightarrow{w_1 \cdot w_2}_p (E \otimes (E_1, E_2), S_1'|S_2')} \\[1em]
(\text{Par4}) & \dfrac{(E, S_1) \xrightarrow{w}_p (E', S_1'), (E, S_2) \not\xrightarrow{}_p}{(E, S_1|S_2) \xrightarrow{w}_p (E', S_1'|S_2)} \\[1em]
(\text{Time}) & \dfrac{(E, S_1) \xrightarrow{\sqrt{}}_n (E_1, S_1'), (E, S_2) \xrightarrow{\sqrt{}}_n (E_2, S_2')}{(E, S_1|S_2) \xrightarrow{\sqrt{}}_n (E \otimes (E_1, E_2), S_1'|S_2')} \\[1em]
(\text{Res}) & \dfrac{(E, S) \xrightarrow{\alpha} (E', S'), \alpha \notin \{a_{\ell,\text{s}}, \overline{a}_{\ell,\text{s}} | a \in L\}}{(E, S\backslash L) \xrightarrow{\alpha} (E', S')\backslash L}
\end{array}
$$

Rules (*Par*1) - (*Par*4) specify how the actions of the components of a parallel composition may be combined. Note that the symmetric versions of these rules are omitted. We point out that, according to rule (*Par*2), if the parallel components may execute complementary actions at the same location, then they may synchronize with each other producing action $\tau_{a,\ell,\mathbf{s}}$. Note that the internal action is decorated by the channel, the location, and the species of the individual that produced an output on the channel during the synchronization. If both components may execute probabilistic transitions then they proceed together with probability the product of the two distinct probabilities (rule (*Par*3)). If exactly one of them enables a probabilistic transition then this transition takes precedence over any non-deterministic transitions of the other component (rule (*Par*4)).

Note that in case that the components proceed simultaneously then the environment of the resulting configuration should take into account the changes applied in both of the constituent transitions (rules (*Par*2), (*Par*3) and (*Time*) as follows:

$$E \otimes (E_1, E_2) = \{(\ell, \mathbf{s}, m + i_1 + i_2) \mid (\ell, \mathbf{s}, m) \in E,$$
$$(\ell, \mathbf{s}, m + i_1) \in E_1, (\ell, \mathbf{s}, m + i_2) \in E_2, i_1, i_2 \in \mathbb{Z}\}$$

Rule (*Time*) defines that parallel processes must synchronize on √ actions. This allows one tick of time to pass and all processes to proceed to their next round. Finally, rule (*Res*) defines the semantics of the restriction operator in the usual way.

**Initial configuration.** Based on this machinery, the semantics of a system $S$ is obtained by applying the semantic rules to the initial configuration. The initial configuration, $(E,S)$, is such that $(\ell,\mathbf{s},m) \in E$ if and only if $S$ contains exactly $m$ individuals of species $\mathbf{s}$ located at $\ell$ of the form $P{:}\langle \mathbf{s},\ell \rangle$, where $P \neq 0$. In general, we say that $E$ is *compatible* with $S$ whenever $(\ell,\mathbf{s},m) \in E$ if and only if $S$ contains exactly $m$ active (non-0) individuals of species $\mathbf{s}$ located at $\ell$. It is possible to prove that the defined semantics preserves compatibility of configurations [3]:

**Lemma 1.** *Whenever* $(E,S) \xrightarrow{\alpha} (E',S')$ *and $E$ is compatible with $S$, then $E'$ is also compatible with $S'$.*

## 2.3 Policies and Prioritized Semantics

We are now ready to define the notion of a policy and refine the semantics of PALPS accordingly. A policy $\sigma$ is a partial order on the set of PALPS non-probabilistic actions. By writing $(\alpha, \beta) \in \sigma$ we imply that action $\beta$ has higher priority than $\alpha$ and whenever there is a choice between $\alpha$ and $\beta$, $\beta$ should always be selected. For example, the policy $\sigma = \{(reproduce_{\ell,\mathbf{s}}, disperse_{\ell,\mathbf{s}}) \mid \ell \in \mathbf{Loc}\}$ specifies that, at each location, dispersal actions of species $\mathbf{s}$ should take place before reproduction actions. On the other hand $\sigma = \{(reproduce_{\ell_1,\mathbf{s}}, disperse_{\ell_1,\mathbf{s}}),(disperse_{\ell_2,\mathbf{s}}, reproduce_{\ell_2,\mathbf{s}})\}$ specifies that, while dispersal should proceed reproduction at location $\ell_1$, the opposite should hold at location $\ell_2$.

To achieve this effect the semantics of PALPS needs to be refined with the use of a new non-deterministic transition system. This new transition relation prunes away all process executions that do not respect the priority ordering defined by the applied policy. Precisely, given a PALPS system $S$ and a policy $\sigma$ then, the semantics of the initial configuration $(E,S)$ under the policy $\sigma$ is given by $\rightarrow_p \cup \rightarrow_\sigma$ where the prioritized nondeterministic transition relation $\rightarrow_\sigma$ is defined by the following rule:

$$\frac{(E,S) \xrightarrow{\alpha}_n (E',S') \text{ and } (E,S) \not\xrightarrow{\beta}_n, (\alpha,\beta) \in \sigma}{(E,S) \xrightarrow{\alpha}_\sigma (E',S')}$$

## 2.4 Examples

*Example 1.* We consider a simplification of the model presented in [42] which studies the reproduction of the parasitic *Varroa mite*. This mite usually attacks honey bees and it has a pronounced impact on the beekeeping industry. In this system, a set of individuals reside on an $n \times n$ lattice of resource sites and go through phases of reproduction and dispersal. Specifically, the studied model considers a population where individuals disperse in space while competing for a location site during their reproduction phase. They produce offspring only if they have exclusive use of a location. After reproduction the offspring disperse and continue indefinitely with the same behavior. In PALPS, we may model the described species $\mathbf{s}$ as $R \stackrel{def}{=} {!}rep.P_0$, where

$$P_0 \stackrel{def}{=} \sum_{\ell \in \mathbf{Nb}(myloc)} \frac{1}{|\mathbf{Nb}(myloc)|} : go\ell.cond(\mathbf{s}@myloc = 1 \triangleright P_1; true \triangleright \sqrt{.}P_0)$$

$$P_1 \stackrel{def}{=} \overline{rep}.(p{:}\sqrt{.}P_0 \oplus (1-p){:}\overline{rep}.\sqrt{.}P_0)$$

We point out that the conditional construct allows us to determine the exclusive use of a location by an individual. The special label *myloc* is used to denote the actual location of an individual within a system definition. Furthermore, note that $P_1$ models the probabilistic production of one or two children of the species. During the dispersal phase, an individual moves to a neighboring location which is chosen equiprobably among the neighbors of its current location. A system that contains two individuals at a location $\ell$ and one at location $\ell'$ can be modeled as

$$System \stackrel{def}{=} (P_0:\langle \ell, \mathbf{s}, 2\rangle | P_0:\langle \ell', \mathbf{s}\rangle | (!rep.P_0):\langle \mathbf{s}\rangle)\backslash\{rep\}.$$

In order to refine the system so that during each cycle of the individuals' lifetime all dispersals take place before the reproductions, we may employ the policy $\{(\tau_{rep,\ell,\mathbf{s}}, \tau_{go,\ell,\mathbf{s}})|\ell,\ell' \in \mathbf{Loc}\}$. Then, according to the PALPS semantics, possible executions of *System* have the form:

$$System \stackrel{w}{\longrightarrow}_p (go\ell_1.\ldots:\langle \ell, \mathbf{s}\rangle | go\ell_2.\ldots:\langle \ell, \mathbf{s}\rangle | go\ell_3.\ldots:\langle \ell', \mathbf{s}\rangle)\backslash\{rep\}$$
$$\stackrel{\tau_{go,\ell_1,\mathbf{s}}}{\longrightarrow}_\sigma (cond(\ldots):\langle \ell_1, \mathbf{s}\rangle | go\ell_2.\ldots:\langle \ell, \mathbf{s}\rangle | go\ell_3.\ldots:\langle \ell', \mathbf{s}\rangle)\backslash\{rep\}$$

for some probability $w$ and locations $\ell_1, \ell_2, \ell_3$, where, in the final state of the above execution, no component will be able to execute the $rep$ action before all components finish executing their movement actions.

*Example 2.* Let us now extend the previous example into a two-species system. In particular, consider a competing species $\mathbf{s'}$ of the Varroa mite, such as the pseudo-scorpion, which preys on $\mathbf{s}$. To model this, we may define the process $R \stackrel{def}{=} !rep'.Q_0$, where

$$Q_0 \stackrel{def}{=} cond(\mathbf{s}@myloc \geq 1 \triangleright Q_1, \mathbf{s}@myloc < 1 \triangleright Q_2)$$
$$Q_1 \stackrel{def}{=} \overline{prey}.Q_3 + \overline{rep'}.Q_4$$
$$Q_2 \stackrel{def}{=} \overline{rep'}.\sqrt{.Q_5}$$
$$Q_3 \stackrel{def}{=} \overline{rep'}.\sqrt{.Q_0}$$
$$Q_4 \stackrel{def}{=} cond(\mathbf{s}@myloc \geq 1 \triangleright \overline{prey}.\sqrt{.Q}, \mathbf{s}@myloc < 1 \triangleright \sqrt{.Q_5})$$
$$Q_5 \stackrel{def}{=} cond(\mathbf{s}@myloc \geq 1 \triangleright \overline{prey}.Q_3, \mathbf{s}@myloc < 1 \triangleright 0)$$

An individual of species $s'$ initially has a choice between preying or producing an offspring. If it succeeds in locating a prey then it preys on it. If it fails then it makes another attempt in the next cycle. If it fails again then it dies.

To implement the possibility of preying on the side of $\mathbf{s}$, its definition must be extended with complementary input actions on channel *prey* at the appropriate places:

$$P_0 \stackrel{def}{=} \sum_{\ell \in \mathbf{Nb}(myloc)} \frac{1}{|\mathbf{Nb}(myloc)|} : (go\ell.cond(\mathbf{s}@myloc = 1 \triangleright P_1; true \triangleright \sqrt{.P_0}) + prey.0)$$
$$P_1 \stackrel{def}{=} \overline{rep}.(p.\sqrt{.P_0} \oplus (1-p)\overline{rep}.\sqrt{.P_0}) + prey.0$$

In this model it is possible to define an ordering between the actions of a single species, between the actions of two different species or even between actions on which individuals of the two different species synchronize. For instance, to specify that preying takes place in each round before individuals of species $\mathbf{s}$ disperse and before individuals of species $\mathbf{s'}$ reproduce we would employ the policy

$$\sigma = \{(\tau_{go,\ell,\mathbf{s}}, \tau_{prey,\ell,\mathbf{s'}}), (\tau_{rep',\ell,\mathbf{s'}}, \tau_{prey,\ell,\mathbf{s'}})|\ell \in \mathbf{Loc}\}.$$

Furthermore, to additionally require that reproduction of species $\mathbf{s}$ precedes reproduction of species $\mathbf{s'}$, we would write $\sigma \cup \{(\tau_{rep',\ell,\mathbf{s'}}, \tau_{rep,\ell,\mathbf{s}})|\ell \in \mathbf{Loc}\}$.

*Example 3.* As a final example, we consider a model inspired by [7] concerning the possible ordering of the activities of reproduction, mortality and dispersal within a single-species individual-based model. In particular, let us assume a species in which individuals may go through reproduction and mortality before dispersing in each cycle of their life. In this species, it is possible to distinguish three cases in which this behavior may take place: reproduction and mortality may take place concurrently within a model, reproduction may proceed mortality for every individual, or reproduction may follow mortality, for every individual. We may model this species in PALPS as follows:

$$P_0 \stackrel{def}{=} mortality.P_m^1 + reproduction.P_r^1$$
$$P_m^1 \stackrel{def}{=} p_m : 0 \oplus (1-p_m) : P_m^2$$
$$P_m^2 \stackrel{def}{=} reproduction.[p_r : \overline{rep_m}.D \oplus (1-p_r) : D]$$
$$P_r^1 \stackrel{def}{=} p_r : \overline{rep_r}.P_r^2 \oplus (1-p_r) : P_r^2$$
$$P_r^2 \stackrel{def}{=} mortality.p_m : 0 \oplus (1-p_m) : D$$
$$D \stackrel{def}{=} \sum_{\ell \in \mathbf{Nb}(myloc)} \frac{1}{|\mathbf{Nb}(myloc)|} : go\ell.\sqrt{.P_0}$$
$$S_1 \stackrel{def}{=} !rep_m.D$$
$$S_2 \stackrel{def}{=} !rep_r.P_r^2$$

According to this definition, an individual of the species may initially nondeterministically select between the activities of mortality ($P_m^1$) and reproduction ($P_r^1$). It then goes through two phases for executing the two activities according to the chosen order, where $p_m$ is the probability of mortality and $p_r$ the probability of reproduction. Note that there are two species processes, namely, $S_1$ and $S_2$. They are distinguished by whether new offspring is exposed to death during the first cycle of their life, as specified in the reproduction before mortality process ordering (process $S_2$).

We may now see that the three orderings discussed above can be implemented via the policies: $\sigma_0 = \emptyset$ for the concurrent ordering,

$$\sigma_1 = \{(\text{reproduction}_{\ell,\mathbf{s}}, \text{mortality}_{\ell,\mathbf{s}}) \mid \ell \in \mathbf{Loc}\}$$

for the mortality-before-reproduction ordering and

$$\sigma_2 = \{(\text{mortality}_{\ell,\mathbf{s}}, \text{reproduction}_{\ell,\mathbf{s}}) \mid \ell \in \mathbf{Loc}\}$$

for the mortality-follows-reproduction ordering. The intuition is that $\sigma_0$ does not impose any order between the two activities, thus, individuals of the species may concurrently engage in reproduction and mortality whereas in $\sigma_1$ and $\sigma_2$ one activity takes priority over another.

## 3 Translating PALPS into PRISM

In this section we turn to the problem of model checking PALPS models extended with policies. As is the case of PALPS without policies, the operational semantics of PALPS *with policies* gives rise to transition systems that can be easily translated to Markov decision processes (MDPs). We recall that Markov decision processes are a type of transition systems that combine probabilistic and non-deterministic behavior. As such, model checking approaches that have been developed for MDPs can also be applied to PALPS models.

PRISM is one such tool developed for the analysis of probabilistic systems. Specifically, it is a probabilistic model checker for Markov decision processes, discrete time Markov chains, and continuous time Markov chains. Using PRISM it is also possible to generate random sample paths of execution for simulation. For our study we are interested in the MDP support of the tool. In [37] we defined a translation of PALPS into the MDP subset of the PRISM language and we explored the possibility of employing the probabilistic model checker PRISM to perform analysis of the semantic models derived from PALPS processes. In this paper, we refine the translation of [37] for taking into account the notion of policies.

In the remainder of this section, we will give a brief presentation of the PRISM language, present an encoding of (a subset of) PALPS *with policies* into PRISM and prove its correctness.

### 3.1 The PRISM language

The PRISM language is a simple, state-based language, based on guarded commands. A PRISM model consists of a set of *modules* which can interact with each other on shared actions following the CSP-style of communication. Each module possesses a set of *local variables* which can be written by the module and read by all modules. In addition, there are *global variables* which can be read and written by all modules. The behavior of a module is described by a set of *guarded commands*. When modeling MDPs these commands take the form:

```
[act] guard   p_1 : u_1 + ... + p_m :u_m;
```

where `act` is an optional action label, guard is a predicate over the set of variables, $p_i \in (0,1]$ and $u_i$ are updates of the form:

```
(x'_1 = u_{i,1}) & ... & (x'_k = u_{i,k})
```

where $u_{i,j}$ is a function over the variables. Intuitively, such an action is enabled in global state $s$ if $s$ satisfies `guard`. If a command is enabled then it may be executed in which case, with probability $p_i$, the update $u_i$ is performed by setting the value of each variable $x_j$ to $u_{i,j}(s)$ (where $x_j'$ denotes the new value of variable $x_j$).

A model is constructed as the parallel composition of a set of modules. The semantics of a complete PRISM model is the parallel composition of all modules using the standard CSP parallel composition. This means that all the modules synchronize over all their common actions (i.e., labels). For a transition arising from synchronization between multiple processes, the associated probability is obtained by multiplying those of each component transition. Whenever, there is a choice of more than one commands, this choice is resolved non-deterministically. We refer the reader to [1] for the full description and the semantics of the PRISM language.

### 3.2 Encoding PALPS *with policies* into the PRISM language

As observed in [29], the main challenge of translating a CCS-like language (like PALPS) into PRISM is how to map binary CCS-style communication over channels to PRISM's multi-way (CSP style) communication. Our approach for dealing with this challenge in [37], similarly to [29], was to introduce a distinct action for each possible binary, channel-based, communication which captures the channel as well as the sender/receiver pair. The two other actions in PALPS, namely the tick action and the movement action, were easily handled via the synchronous communication of

PRISM, in the case of the tick action, and via a single PRISM command, in the case of the movement action.

In PALPS *with policies* the translation becomes more complex because, at any point, we need to select actions that are not preempted by other enabled actions. For example, suppose that according to our policy $\sigma$, $(\beta) \in \sigma$. This implies that, at any point during computation, we must have information as to whether $\beta$ is enabled. To implement this in PRISM, we employ a variable $n_\beta$ which records the number $\beta$'s enabled. To begin with, this variable is initialized with the relevant values as given rise to by the model. Subsequently, it is updated as computation proceeds: once a $\beta$ is executed then $n_\beta$ is decreased by 1 and when a new occurrence becomes enabled it is increased by 1. For example, given the process $\eta_1.P_1 + \eta_2.P_2$, if action $\eta_1$ is executed and if additionally $\eta_2 = \beta$, then $n_\beta$ should be decreased by 1 and, if $P_1$ enables $\beta$ then $n_\beta$ should be increased accordingly. Finally, we point out that, if $(\beta) \in \sigma$, execution of action $\$ in any module of a model should have as a precondition that $n\_\beta = 0$.

To translate PALPS into the PRISM language, we translate each process into a module. The execution flow of a process is captured with the use of a local variable within the module whose value is updated in every command in such as way as to guide the computation through the states of the process. Then, each possible construct of PALPS is modeled via a set of commands. For example, the probabilistic summation is represented by encoding the probabilistic choices into a PRISM guarded command. Non-deterministic choices are encoded by a set of simultaneously enabled guarded commands that capture all nondeterministic alternatives, whereas the conditional statement is modeled as a set of guarded commands, where the guard of each command is determined by the expressions of the conditional process.

Unfortunately, the replication operator cannot be directly encoded into PRISM since the PRISM language does not support the dynamic creation of modules. To overcome this problem, we consider a bounded replication construct of the form $!^m P$ in which we specify the maximum number of $P$'s, namely $m$, that can be created during computation. We note that, in practice, the value of $m$ can be selected by estimating a bound of the maximum size of the population, or it can be determined by the size of the state-space of the resulting model.

We now consider the main ideas of translating PALPS into the PRISM language via an example.

*Example 4.* Consider a habitat consisting of four patches $\{1,2,3,4\}$, where **Nb** is the symmetric closure of the set $\{(1,2),(1,3),(2,4),(3,4)\}$. Let **s** be a species residing on this habitat defined according to the bounded replication $R$:

$$\begin{aligned}
R &\stackrel{def}{=} !^m rep.P_1 \\
P_1 &\stackrel{def}{=} disperse.P_2 + reproduce.P_3 \\
P_2 &\stackrel{def}{=} \sum_{\ell \in \mathbf{Nb}(myloc)} \frac{1}{2} : go\ell.P_4 \\
P_3 &\stackrel{def}{=} p : \overline{rep}.P_5 \oplus (1-p) : P_5 \\
P_4 &\stackrel{def}{=} p : \overline{rep}.\sqrt{.P_1} \oplus (1-p) : \sqrt{.P_1} \\
P_5 &\stackrel{def}{=} \sum_{\ell \in \mathbf{Nb}(myloc)} \frac{1}{2} : go\ell.\sqrt{.P_1}
\end{aligned}$$

According to the definition, an individual of species **s** begins by nondeterministically selecting between the activities of dispersal and reproduction. If dispersal it selected, then it migrates with equal probability to one of the neighboring locations and then probabilistically produces an offspring before returning to its initial state. If reproduction is selected, then the order between these two activities is swapped. Now, consider a system initially consisting of two individuals, one at location 1 and one at location 4:

$$System \stackrel{def}{=} (P_1:\langle \mathbf{s},1\rangle \mid P_1:\langle \mathbf{s},4\rangle \mid R:\langle \mathbf{s}\rangle) \backslash \{rep\}$$

Further, suppose that we would like to analyze the system under the policy

$$\sigma = \{(\text{reproduce}_{\ell,\mathbf{s}}, \text{disperse}_{\ell,\mathbf{s}}), (\tau_{rep,\ell,\mathbf{s}}, \tau_{go,\ell,\mathbf{s}}) \mid \ell \in \mathbf{Loc}\}.$$

That is, we are interested in a process ordering where dispersal takes place before reproduction and all movement actions proceed the reproduction synchronizations.

In order to translate *System* under policy $\sigma$ in the PRISM language we first need to encode global information relating to the system. This consists of four global variables that record the initial populations of each of the locations and two variables that record the number of enabled occurrences of the actions of the higher priority referred to in the policy $\sigma$, that is, of disperse$_\mathbf{s}$ and $\tau_{go,\mathbf{s}}$. We also include a global variable $i$ that measures the inactive individuals still available to be triggered. Initially $i = m$.

```
global s1, s4: [0,m+2] init 1;
global s2, s3: [0,m+2] init 0;
global i: [0,m] init m;
global n_d: [0,m+2] init 2; //dispersal action
global n_g: [0,m+2] init 0; //go action
```

We continue to model the two individuals $P_1:\langle \mathbf{s},1\rangle$ and $P_1:\langle \mathbf{s},4\rangle$. Each individual will be described by a module. In Fig. 1, we may see the translation of individual $P_1:\langle \mathbf{s},1\rangle$.

```
module P1

st1 : [1..15] init 1;
loc1: [1..4] init 1;
const int s1 = 1;

[disperse] (st1=1)&(pact=0) -> (st1'=2)&(n_d'=n_d-1)&(pact'=1);
[reproduce] (st1=1)&(n_d=0)&(pact=0) -> (st1'=3)&(n_d'=n_d-1);

[prob] (st1=2) -> 0.5:(st1'=6)&(n_g'=n_g+1)&(pact'=0)
              + 0.5:(st1'=7)&(n_g'=n_g+1)&(pact'=0);
[] (st1=6)&(loc=1)&(pact=0) -> (loc'=2)&(s1'=s1-1)&(s2'=s2+1)
                          &(n_g'=n_g-1)&(st'=4);
[] (st1=7)&(loc=1)&(pact=0) -> (loc'=3)&(s1'=s1-1)&(s3'=s3+1)
                          &(n_g'=n_g-1)&(st'=4);
... // All possible moves are enumerated

[prob] (st1=4) -> 0.5:(st1'=8) + 0.5:(st1'=13)&(pact'=0);

[] (st1=8)&(i>0)&(n_g=0)&(pact=0) -> (s1'=s1+1)&(i'=i-1)&(st'=9);
[rep_1_3] (st1=9)&(pact=0) -> (st1'=13); // Activate module 3
... // All activation possibilities are enumerated

[prob] (st1=3) -> 0.5:(st1'=10)&(pact'=0)
              + 0.5:(st1'=5)&(pact'=0);

[] (st1=10)&(i>0)&(n_g=0)&(pact=0) ->
                         (s1'=s1+1)&(i'=i-1)&(st'=11);
[rep_1_3] (st1=11)&(pact=0) -> (st1'=5)&(pact'=1);
... // All activation possibilities are enumerated

[prob] (st1=5) -> 0.5:(st1'=11)&(n_g'=n_g+1)&(pact'=0)
              + 0.5:(st1'=12)&(n_g'=n_g+1)&(pact'=0);
[] (st1=11)&(loc=1)&(pact=0) -> (loc'=2)&(s1'=s1-1)&(s2'=s2+1)
                          &(n_g'=n_g-1)&(st'=13);
[] (st1=12)&(loc=1)&(pact=0) -> (loc'=3)&(s1'=s1-1)&(s3'=s3+1)
                          &(n_g'=n_g-1)&(st'=13);
... // All possible moves are enumerated

[tick] (st1=13) -> (st1'=14);
[] (st1=14) -> (n_d'=n_d+1)&(st1'=15);
[tick'] (st1=15) -> (st1'=1);

[prob] (pact=1)&(st1!=2,3,4,5) -> (pact'=0);
endmodule
```

**Fig. 1:** PRISM code for an active individual

We observe, that its species variable, $s1$, is set to 1, a constant that identifies the species, its location variable, $loc1$, is set to 1 and variable $st1$, recording its state, is set to 1, the initial state of the module. Overall, the module has 15 different states. From state 1 two commands are enabled: one of actions disperse and reproduce can take place, though the latter has as a precondition that $n_d = 0$. Thus, in fact, it will never be enabled. Then from state 2, a probabilistic transition takes place to determine the position of dispersal. This yields one of the states 6 and 7 which result in horizontal and vertical dispersals along the grid, respectively. In both bases variable $n_g$ is increased by one as the *go* action becomes enabled. Note that the actions enabled from states 6 and 7 update the number of individuals of the source and destination locations of the move and the variable $n_g$ is decreased by one as there is now one fewer movement action enabled. From state 4 a probabilistic transition determines whether the individual will reproduce or not. In the case that dispersal is selected, it is executed in two distinct steps: initially at state 8 it is confirmed that there is still an individual to activate and that no movement actions are currently enabled. In this case variables $i$ and $s_{loc}$ are updated and the flow of control is passed on to state 9 where a synchronization with an inactive module is performed. Finally, we point out that the tick action is implemented via three actions in PRISM (states 13-15): initially all modules are required to synchronize on the *tick* action, then they all perform their necessary updates for actions that will become enabled by the move and, finally, the modules are required to synchronize again before they may start to execute their next time step.

Note that for both the reproduction and the tick actions the moves given rise by our translation cannot be merged into one due to the restriction of PRISM that commands which synchronize with other modules (such as *rep*) cannot modify global variables (such as $s_i$).

Individual $P_1$:⟨**s**,4⟩ may be defined similarly. Note that the module encoding this process is identical to module $P1$ with the exception that we rename the names of the variables and the initial value of the location variable.

This leaves us with the encoding of $R$: the component that implements replication of individuals. As we have already discussed, we achieve this via bounded replication which makes an assumption on the maximum number of new individuals that can be created in a system. Given this assumption, our model must be extended by an appropriate number of inactive individuals awaiting for a trigger via a *rep_i_j* action as illustrated in Fig. 2.

```
module P3
```

```
st3 : [0..15] init 0;
loc3: [1..4] init 1;
const int s3 = 1;
[tick] (st3=0) -> (st3'=0);
[tick'] (st3=0) -> (st3'=0);
[rep_1_3] (st3=0) -> (st3'=1)&(loc3=loc1); ...
[rep_max_3] (st3=0) -> (st3'=1)&(loc3=loc_max); ...
// Here we append the code of an active $P_1$ individual
// with the variables appropriately renamed

endmodule
```

**Fig. 2:** PRISM code for an inactive individual

---

Thus, the inactive individual modeled by module P3, awaits to synchronize with any of the remaining modules $0,...,m + 2$, in which case it inherits the location of the synchronizing module and it sets $st3 = 1$ so that it may begin to execute the code of an active individual, presented in Fig. 1, with the variables appropriately renamed.

### 3.3 Formal translation

In this section, we will formalize the intuitions of the previous example into a formal translation of PALPS into PRISM and we will prove its correctness.

Consider a PALPS model. This consists of a set of locations, $L = \{1,...,k\}$, a set of attributes, $\Theta = \{\theta_1,...,\theta_m\}$, and a value of each attribute at each location, the neighborhood relation **Nb**, a process *System* and a policy $\sigma = \{(1,\beta_1), ...,(p,\beta_p)\}$. We assume, for the reasons already discussed in the previous section, that all replication processes are bounded and have the form $R =!^n rep.P$, thus allowing the creation of up to $n$ individuals of the specific species. We also, assume that all *rep* channels are restricted within *System*. Then, the PRISM model is constructed as follows:

- For each species **s**, the model contains the $k$ global integer variables $s_1,...,s_k$, capturing the number of individuals of species **s** for each of the locations. The variables are appropriately initialized based on the definition of *System*.
- For each attribute $\theta$ and each location $\ell$, the model contains a constant that records the value of $\theta_\ell$.
- For each channel $a$ on which synchronization takes place we introduce a variable $a_y$ which counts the number of available inputs at location $y$.
- There exists a global variable *atomic* which may take values from $\{0,1\}$ and is used to force the atomic execution of sequences of actions forming the translation of a single PALPS action. Initially, *atomic* = 0.
- There exists a global variable *pact* which may take values from $\{0,1\}$ and expresses whether there is a probabilistic action enabled. It is used to give precedence to probabilistic actions over nondeterministic actions. Initially, *pact* = 0. Furthermore, all non-probabilistic actions have *pact* = 0 as a precondition.
- For each action $\beta$ such that $(\beta) \in \sigma$, we distinguish two cases. If $\beta$ is an input/output action on a channel or the action $\tau_{go,\ell,\mathbf{s}}$, then we introduce a variable $n_\beta$ which counts the number of enabled actions of type $\beta$. If instead $\beta = \tau_{a,\ell,\mathbf{s}}$ for some channel $a$, then we employ three variables $n_{a,\ell}$, $n_{\overline{a},\ell}$ and $n_\beta$ which count the available occurrences of $a_{\ell,\mathbf{s}}$, $\overline{a}_{\ell,\mathbf{s}}$ and $\beta$, respectively.
- Each (active) process $P:\langle\mathbf{s},\ell\rangle$ of *System* becomes a PRISM module with a constant $sp_P = \mathbf{s}$, a variable $loc_P = \ell$ and a variable $st_P$, with range $1,...,|P|$, which records the current state of the individual and where $|P|$ is the number of states that $P$ may evolve to. The body of the module is the translation of process $P$ into guarded commands.
- Each species definition $R:\langle\mathbf{s}\rangle =!^n rep.P$ of *System* becomes a sequence of $n$ PRISM modules, $P_{x+1},...,P_{x+n}$, where $x$ is the number of individuals of species **s** in the initial state of *System*. In our model, we introduce a variable $i_\mathbf{s}$ that records the current number of inactive individuals of species **s**:

$$i_s : [0,n] \text{ init } n;$$

Each inactive module $P_y$ possesses a constant $sp_y = \mathbf{s}$, a variable $loc_y$ which is not initialized, and another variable $st_y$, with range $0,...,|P|$, which corresponds to the current state of the individual and where $|P|$ is the number of states that $P$ may evolve to. Note that $st_y = 0$ corresponds to the state where the inactive individual is awaiting activation by one of the active individuals of species **s**. To capture this, we include the following commands:

```
module P_y
st_y : [0..|P|] init 0;
loc_y : [1..m];
const int s_y = s;
[tick](st_y = 0) ⟶ (st'_y = 0)
[tick'](st_y = 0) ⟶ (st'_y = 0)
[rep_{1,y}](st_y = 0) ⟶ (st'_y = 1) & (loc_y = loc_1);
...
[rep_{x+n,y}](st_y = 0) ⟶ (st'_y = 1) & (loc_y = loc_x);
...
//Here we append the translation of P
endmodule
```

Thus, $P_y$ may be activated by any of the modules $P_1,…,P_{x+n}$ and then proceed according to the translation of process $P$ into guarded commands.

We now continue to describe how a process at the individual level of PALPS can be translated to a sequence of PRISM commands. We denote the translation of a process $P$ as [ [$P$] ] and we define it inductively on the structure of $P$. Note that, for convenience, we write $s_Q$ to for the state (integer value) associated with process $Q$. In the translations below, we assume that we are working within a module with identifier $x$, species **s**, and variables $st$, and $loc$. Furthermore, we write

- enabled($P$) for the set of all input actions $a_\ell$ such that $P \xrightarrow{a_\ell}$,
- enabled$_\sigma$($P$) for the set of all actions such that, for all ($\beta$, ∈ $\sigma$, if ≠ $\tau_{a,\ell}$ then ∈ enabled$_\sigma$($P$), and if $\bar{\tau}_{a,\ell}$ then $a_\ell, \bar{a}_\ell$ ∈ enabled$_\sigma$($P$),
- enabled$_{\sigma,\tau}$($P$) for the set of all actions $\bar{\tau}_{a,\ell,\mathbf{s}}$ such that $P \xrightarrow{\alpha}$ and ($\beta$, ∈ $\sigma$, and
- prob($P$) for the logical value that expresses whether $P$ is a probabilistic process.

Based on these notions, we write

$$\text{updates}(P, Q) = \bigwedge_{a_\ell \in \text{enabled}(Q)} (a'_\ell = a_\ell + 1) \& \bigwedge_{a_\ell \in \text{enabled}(P)} (a'_\ell = a_\ell - 1)$$
$$\& \bigwedge_{\mu \in \text{enabled}_\sigma(Q)} (n'_\mu = n_\mu + 1) \& \bigwedge_{\mu \in \text{enabled}(P)} (n'_\mu = n_\mu - 1)$$
$$\& \bigwedge_{\tau_{a,\ell} \in \text{enabled}_{\sigma(Q),\tau}} n'_{\tau_{a,\ell}} = \min(n_{a_\ell}, n_{\bar{a}_\ell}) \& pact = \text{prob}(P)$$

Finally, in the translations below, we assume that we are working within a module with identifier $x$ and variables $st$, and $loc$.

*Case 1: $Q$ = go l.P.* We translate the process by including the command

$$[] \quad (st = s_Q) \& (loc = a) \& (atomic = 0) \& (pact = 0) \& ((a, \ell) \in \mathbf{Nb}) \& (\mathbf{n}_\beta = \mathbf{0})$$
$$\longrightarrow (st' = s_P) \& (loc' = l) \& (st'_a = st_a - 1) \& (st'_l = st_l + 1) \& \text{updates}(Q, P);$$

where the highlighted condition ($\mathbf{n}_\beta = \mathbf{0}$) is only present if there exists ($\tau_{go,\ell},\beta$) ∈ $\sigma$. We then append the translation of $P$. Note that according to its definition, updates($Q,P$) will increase by 1 all variables $n$

  *wheretheaction*$ is enabled by $Q$ while reducing $n$

,$\bar{\tau}_{go,\ell,\mathbf{s}}$ by 1, since there is now one less occurrence of the specific action.

*Case 2: $Q$ = a.P.* To begin with, the process is translated into the following command which captures the possibility that $Q$ executes the input action on channel $a$ independently of synchronizing output actions.

$$[a_x](st = s_Q) \& (atomic = 0) \& (pact = 0) \& (\mathbf{n}_\beta = \mathbf{0}) \longrightarrow (st' = s_P) \& \text{updates}(Q, P);$$

where the highlighted condition is included only if there exists ($a_{loc,\mathbf{s}},\beta$) ∈ $\sigma$. Additionally, we include the commands

$$[a_{y,x}] \quad (st = s_Q) \& (loc = loc_y) \longrightarrow (st' = s_{Q_1});$$
$$[] \quad (st = s_{Q_1}) \longrightarrow (st' = s_P) \& \text{updates}(Q, P) \& (atomic' = 0);$$

for each module $y$ that may perform an output on channel $a$. Note that this transition does not have the requirement that *atomic* = 0. This will be explained in conjunction with the translation of the action $\bar{a}$ in Case 3, below. We then append the translation of $P$.

*Case 3: $Q$ = $\bar{a}.P$.* Similarly to Case 2, we include the following command to capture that $Q$ may execute $\bar{a}$ independently of any synchronizing action on channel $a$.

$$[a'_x] \quad (st = s_Q) \& (atomic = 0) \& (pact = 0) \& (\mathbf{n}_\beta = \mathbf{0}) \longrightarrow (st' = s_P) \& \text{updates}(Q, P);$$

where the highlighted condition is included only if there exists ($\bar{a}_{loc,\mathbf{s}},\beta$) ∈ $\sigma$.

Furthermore, we include the commands

$$[\,]\ (st = s_Q)\&(a_{loc} > 0)\&(atomic = 0)\&(pact = 0)\&(\mathbf{n}_\beta = \mathbf{0}) \longrightarrow$$
$$(st' = s_{Q_1})\&updates(Q,P)\&(atomic = 1);$$
$$[a_{x,y}]\ (st = s_{Q_1})\&(loc = loc_y) \longrightarrow (st' = s_P);$$

for each module $y$ that may perform an input on channel $a$. Note that in this piece of code, initially a process willing to perform an output checks whether there is another process willing to do an input at the same location ($a_{loc} > 0$), assuming that *atomic* = 0 and *pact* = 0. In this case, it performs all its updates and sets *atomic* = 1. It then proceeds to synchronize with a process ready to do an input via action $a_{x,y}$ at its location. Note that such an action is enabled as an input without a restriction of *atomic* = 0. We may then observe that the process synchronizing on the input (see Case 2) will continue to perform via a second action its own updates and set *atomic* = 0. We point out that splitting this synchronization in three distinct commands was necessary due to the fact that in PRISM, actions in which synchronization is performed (such as $a_{x,y}$) do not permit to update global variables as necessary by updates($Q,P$).

As before the condition ($n_\beta = 0$) is included only if there exists $(\tau_{\ell,\mathbf{s}},\beta) \in \sigma$. We then append the translation of $P$.

*Case 4: $Q = \sqrt{.P}$.* We translate the process by including the commands

$$[tick](st = s_Q) \longrightarrow (st' = s_{Q_1});$$
$$[\,](st = s_{Q_1}) \longrightarrow updates(Q,P)\&(st' = s_{Q_2});$$
$$[tick'](st = s_{Q_2}) \longrightarrow (st' = s_P);$$

and appending the translation of $P$. Note that, as in the case of a synchronization, this action needs to be split in three steps: in the first step all modules synchronize on the *tick* action, they each then perform their updates, and, before any module may proceed, the modules are forced to synchronize on the *tick'* action. This is again necessary since the necessary updates cannot be performed while the module are synchronizing on their *tick* actions.

*Case 5: $Q = rep.P$.* We translate the process by including the commands

$$[rep_{y,x}]\ (st = s_Q) \longrightarrow (st' = s_{Q_1});$$
$$[\,]\ (st = s_{Q_1}) \longrightarrow (st' = s_P)\&updates(Q,P)\&atomic' = 0;$$

for each module $y$ that may perform an output on channel *rep*. We then append the translation of $P$. The translation is explained in conjunction with the next case.

*Case 6: $Q = \overline{rep}.P$.* We translate the process by including the commands

$$[\,]\ (st = s_Q)\&(i_s > 0)\&(atomic = 0)\&(pact = 0)\&(\mathbf{n}_\beta = \mathbf{0}) \longrightarrow$$
$$(i'_s = i_s - 1)\&(st' = st_{Q_1})\&updates(Q,P)\&(atomic' = 1);$$
$$[rep_{x,y}]\ (st = s_{Q_1}) \longrightarrow (st' = s_P);$$

for each inactive module $y$ of species $\mathbf{s}$, and then appending the translation of $P$. As before, the highlighted condition is included only if there exists $(\tau_{rep,\ell,\mathbf{s}},\beta) \in \sigma$. Note that in this case, a module aiming to perform the reproduction action $\overline{rep}$ begins by confirming that there are inactive modules of its species still available ($i_s > 0$) and that *atomic* = 0 and *pact* = 0. In such a case, it reduces $i_s$ by 1, it performs its updates and it sets *atomic'* = 1 so that the next two steps are performed atomically. These steps consists of a synchronization of the current module with the module of an inactivated individual (action $rep_{x,y}$) after which the newly-activated individual will perform its updates and set *atomic* = 0.

*Case 7: $Q = 1.P_1 + 2.P_2$.* We translate the process by computing the translations [ [$1.P_1$] ] and [ [$2.P_2$] ] and replacing all commands of the form

$$[act](st = st_{\alpha_1.P_1})\ guard \longrightarrow updates;$$

by

$$[act](st = st_Q)\ guard \longrightarrow updates';$$

where *updates'* is the same as *updates* expect that we compute updates($Q,P_1$) instead of updates($1.P_1,P_1$), and similarly for the commands of [ [$2.P_2$] ].

*Case 8: $Q = p_1 : P_1 + \ldots + p_n : P_n$.* We translate the process by appending [ [$P_1$] ],...,[ [$P_n$] ] to the command:

$$[prob](st = s_Q) \longrightarrow p_1 : (st' = s_{P_1})\&updates(Q,P_1) + \ldots$$
$$+ p_n : (st' = s_{P_n})\&updates(Q,P_n);$$

*Case 9: $Q = cond\ (e_1 \trianglelefteq P_1,\ldots,e_n \trianglelefteq P_n)$.* We translate the process by constructing [ [$P_1$] ],...,[ [$P_n$] ] and replacing each command of the form:

$$[act](st = s_{P_i})\&guard \longrightarrow updates;$$

by the command

$$[act](st = s_Q)\&![[e_1@loc]]\&\ldots\&![[e_{(i-1)}@loc]]\&[[e_i@loc]]\&guard \longrightarrow updates;$$

where [ [$e@loc$] is the translation of the PALPS expression $e@\ell$ into the PRISM language.

*Case 10: $Q = P\backslash L$.* We translate the process by computing [ [$P$] ] and then removing all transitions with label [$a_i$] and [$a_i'$] where $a \in L$.

*Case 11 $C$, $C \stackrel{def}{=} P$.* We translate the process by computing [ [$P$] ] and replacing each command in [ [$P$] ] of the form

$$[\,](st = s_P)\&guard \longrightarrow updates;$$

by

$$[\,](st = s_Q)\&guard \longrightarrow updates;$$

*Case 11: $Q = 0$.* We translate the process as

$$[\,](st = s_0)\&(loc = l) \longrightarrow (s'_t = s_t - 1)\&(st' = s_{done});$$
$$[tick](st = s_{done}) \longrightarrow (st = s_{done});$$
$$[tick'](st = s_{done}) \longrightarrow (st = s_{done});$$

Finally, in each module we include a transition that allows the module to perform a probabilistic action, assuming that $pact = 1$, i.e. there exists at least one module willing to execute a probabilistic action. In this way probabilistic actions are given precedence over nondeterministic actions as required by the semantics. Note that a module resorts to this action only if itself does not enable a probabilistic action. Assuming that the probabilistic states of the module are $s_1,\ldots,s_p$, this remaining action is as follows:

$$[prob](pact = 1)\&(st! = s_1)\&\ldots\&(st! = s_p) \longrightarrow (st = s_{done});$$

### 3.4 Correctness of the translation

We now turn to consider the correctness of the proposed translation. This is demonstrated via the following two theorems. In what follows, given a PRISM model $M$, we write $M \xrightarrow{\alpha,p_i} M_i$ if $M$ contains an action `[α] guard -> p_1 : u_1 + ... + p_m : u_m;` where guard is satisfied in model $M$ and execution of $u_i$ gives rise to model $M_i$. Furthermore, we write $M \xrightarrow{\alpha,1}{}^m M'$ if $M(\xrightarrow{\alpha,1})^m M'$, that is, $M$ may evolve into $M'$ after an a sequence of $m$ moves each of which is executed with probability 1.

**Theorem 1.** For any configuration $(E,Sys)$ and policy $\sigma$, where $E$ is compatible with $Sys$, the following hold:

1. if $(E,Sys) \xrightarrow{\mu}_\sigma (E',Sys')$ then $[\,[(E,Sys)]\,] \xrightarrow{}{}^m [\,[(E',Sys')]\,]$, for some $m$,
2. if $(E,Sys) \xrightarrow{w}_p (E',Sys')$ then $[\,[(E,Sys)]\,] \xrightarrow{prob,w} [\,[(E',Sys')]\,]$.

**Theorem 2.** For any configuration $(E,Sys)$ and policy $\sigma$, where $E$ is compatible with $Sys$, the following hold:

1. if $[\,[(E,Sys)]\,] \xrightarrow{prob,w} M$ then $(E,Sys) \xrightarrow{w}_p (E',Sys')$ and $M = [\,[(E',Sys')]\,]$,
2. if $[\,[(E,Sys)]\,] \xrightarrow{a,1} M$, then $(E,Sys) \xrightarrow{a}_\sigma (E',Sys')$ and $M \xrightarrow{}{}^m M'$ for some $m$, where $M' = [\,[(E',Sys')]\,]$ and whenever $M \xrightarrow{}{}^m M''$ then $M'' = [\,[(E',Sys')]\,]$.

Theorem 1 establishes that each transition of $(E,Sys)$ can be mimicked by its translation module in a sequence of steps: in the case of probabilistic actions this is achieved in a single step, whereas in the case of nondeterministic actions, this may take more than one step in the PRISM translation. Theorem 2 considers the other direction of the correctness: Given a transition of a PRISM module there are two possibilities. If the transition is a *prob* transition, when a probabilistic action with the same probability may take place at the PALPS level. Otherwise, it is possible that the transition of the module has resulted in the first of a sequence of states for establishing the transition of a PALPS process. In this case, the intermediate state may perform no other execution steps other than to reach the translation of the resulting process of the PALPS process.

**Sketch of the proof of Theorem 1:** The proof consists of a case analysis of all possible ways in which the transition $(E,Sys) \xrightarrow{\alpha} (E',Sys')$ can be produced. Four cases exist:

- If the transition involves a single process participant $P:\langle \mathbf{s},\ell \rangle$, then we may verify by induction on the structure of $P$ that any action $P$ can perform can also be performed by its translation and the resulting PRISM model is the translation of $(E',Sys')$. In all cases this can be established in a single move of module $P$.
- If instead the transition has arisen via the communication of two components of $Sys$ then it is possible to establish that the two modules corresponding to the two components share the action in question and can thus execute the synchronization. This will take three actions at the PRISM level.
- If $\overline{\sqrt{}}$, then it must be that all components of $Sys$ enable the transition $\sqrt{}$. We may then observe that the PRISM translations of the components enable the *tick* action and thus the transition can be performed in a sequence of moves.
- if $\overline{w}$, then it must be that a set of $Sys$ processes enable a probabilistic transition and $w$ is the product of the associated probabilities. We may then observe that all PRISM components enable the *prob* action with the respective modules enabling the specific probabilistic actions and the remaining modules enabling the action with probability 1. As a result the PRISM model will match the transition with a $(prob,q)$ action. The resulting PRISM model is the translation of $(E',Sys')$. □

**Sketch of the proof of Theorem 2:** The proof consists of a case analysis of all possible ways in which the transition $[\,[(E,Sys)]\,] \xrightarrow{(a,p)} M$ can be produced. It follows along similar lines with the proof of Theorem 1. The interesting cases include the synchronizations, the activations of inactive modules, and the *tick* action. The important point to note here is that, in all cases, the intermediate step $M$ captures correctly environment $E'$ in the transition $(E,Sys) \xrightarrow{\alpha} (E',Sys')$. Furthermore, the assignment *atomic* = 1 locks all actions not involved in the completion of the translation of $(E,Sys) \xrightarrow{\alpha} (E',Sys')$. As result, there exists exactly one possible path of execution of the PRISM model which is exactly the one leading to $[\,[(E',Sys')]\,]$. □

### 3.5 Verification in PRISM

In this section we briefly describe the types of analysis that can be performed on PALPS models via the PRISM model checker.

**Model Checking** To begin with, PALPS models may be model checked in PRISM against properties specified in the PCTL logic [10]. The syntax of the PCTL logic is given by the following grammar where $\Phi$ and $\phi$ range over PCTL state and path formulas, respectively, $p \in [0,1]$ and $k \in \mathbb{N}$.

$$\Phi := true \mid e \mid \neg\Phi \mid \Phi \wedge \Phi' \mid P_{\bowtie p}[\phi]$$
$$\phi := X\Phi \mid \Phi U^k \Phi \mid \Phi_1 U \Phi_2$$

In the syntax above, we distinguish between state formulas $\Phi$ and path formulas $\phi$, which are evaluated over states and paths, respectively. A state formula is built over logical expressions $e$ and the construct $P_{\bowtie p}[\phi]$. Intuitively, a configuration $s$ satisfies property $P_{\bowtie p}[\phi]$ if for any possible execution beginning at the configuration, the probability of taking a path that satisfies the path formula $\phi$ satisfies the condition $\bowtie p$.

Path formulas include the $X$ (next), $U^k$ (bounded until) and $U$ (until) operators, which are standard in temporal logics. Intuitively, $X\Phi$ is satisfied in a path if the next state satisfies path formula $\Phi$. Formula $\Phi_1 U^k \Phi_2$ is satisfied in a path if $\Phi_1$ is satisfied continuously on the path until $\Phi_2$ becomes true within $k$ time units (where time units are measured by $\sqrt{}$ events in PALPS). Finally, formula $\Phi_1 U \Phi_2$ is satisfied if $\Phi_2$ is satisfied at some point in the future and $\Phi_1$ holds up until then.

As an example, consider a population **s** in danger of extinction. A property that one might want to check for such a population is that the probability of extinction of the population in the next ten years is less than a certain threshold $p_e$. This can be expressed in PCTL by the property $P_{\leq p_e}[trueU^{10} \sum_{\ell \in \mathbf{Loc}} \mathbf{s}@\ell = 0]$. Alternatively, one might express that a certain central location $\ell$ will be re-inhabited with at least some probability $p_r$ by $\mathbf{s}@\ell = 0 \rightarrow P_{\geq p_r}[trueU(\mathbf{s}@\ell > 0)]$.

It is also possible to study the relation within a model between the size of the initial population and the probability of extinction of the population, by checking properties of the form $\mathbf{s}@\ell \geq m \rightarrow P_{\geq p_r}[trueU(\mathbf{s}@\ell = 0)]$ or to explore the dynamics between two (or more) competing populations **s** and **s**′. As an example, expressing that, within the next 20 years with some high probability, members of the population **s** will outnumber the members of population **s**′: $P_{\geq p}[trueU(\sum_{\ell \in \mathbf{Loc}} \mathbf{s}'@\ell \leq \sum_{\ell \in \mathbf{Loc}} \mathbf{s}@\ell)]$.

PRISM also enables to take a more *quantitative* approach for model checking PCTL properties: it supports the verification of the constructs $P_{min=?}[\phi]$ and $P_{max=?}[\phi]$ via which the minimum and maximum probabilistic of satisfying $\phi$ are computed.

**Steady-state behavior.** PRISM also supports reasoning about the *steady-state behavior* of a model, that is, the behavior in the long-run or when an equilibrium is reached [10]. Steady-state properties are only available for discrete-time and continuous-time Markov chains. These properties are expressed by $S_{bound}[prop]$. Such a property is true in a state $s$ of a discrete-time or a continuous-time Markov chain if, starting from $s$, the steady-state (long-run) probability of being in a state which satisfies the (boolean-valued) property *prop*, meets the bound *bound*. For example, the steady-state property $S_{bound}[s@2 = 4]$ expresses that the long-run probability that there will be 4 individuals of species $s$ at location 2 meets the bound.

**Rewards.** PRISM models can also be augmented with information about rewards: It is possible to assign a reward (a positive real number) to any command or state of a PRISM model. Every time a command is executed or a state is visited, the rewards associate with the command or state is accumulated. It is then possible to reason about reward-based properties for discrete-time Markov chains, by extending the logic PCTL with the following additional operators [1]:

$$R_{\bowtie r}[C^{\leq k}] \mid R_{\bowtie r}[I^{\leq k}] \mid R_{\bowtie r}[F\Phi] \mid R_{\bowtie r}[S]$$

where $\bowtie \in \{<, \leq, \geq, >\}, r \in \mathbb{R}_{\geq 0}, k \in \mathbb{N}$ and $\Phi$ is a PCTL formula. The $R$ operator defines properties about the *expected* value of rewards. The formula $R_{\bowtie r}[\psi]$, where $\psi$ denotes one of the four possible operators in the grammar above, is satisfied in a state $s$ if, from $s$, the expected value of reward $\psi$ meets the bound $\bowtie r$. Operator $C^{\leq k}$ refers to the reward accumulated over k time steps; $I^{\leq k}$ the state reward at time instant $k$; $F\Phi$, the reward accumulated before a state satisfying $\Phi$ is reached; and $S$, the long-run rate of reward accumulation. Properties of the form $R_{=?}[\psi]$ means "what is the expected reward for operator $\psi$?".

## 4 A case study in PRISM

In this section, we apply our methodology for the simulation and model checking of PALPS systems using the PRISM tool. As a case study we consider a variation of the system in Example 1, Section 2.4, which was also considered in [37] and can thus serve as a benchmark for studying the effect of applying policies on systems and, in particular the degree by which policies reduce the state space of a PRISM model. The variation we hereby consider, as we can see below, is that the order of dispersal and reproduction is not fixed: the two activities can take place in an arbitrary order.

In our model we will assume a lattice of locations of size $n \times n$ (we will consider $n = 4, 9, 16$). We assume periodic boundaries conditions so that the opposite sides of the grid are connected together. Then, the PALPS definition of an individual takes the following form:

$$P \stackrel{def}{=} \text{disperse}.P_1 + \text{reproduce}.P_2$$
$$P_1 \stackrel{def}{=} \sum_{\ell \in \mathbf{Nb}(myloc)} \frac{1}{4} : go\ell.cond(\mathbf{s}@myloc = 1 \triangleright P_3; true \triangleright \sqrt{.P})$$
$$P_3 \stackrel{def}{=} \overline{rep}.(p : \sqrt{.P} \oplus (1-p) : \overline{rep}.\sqrt{.P})$$
$$P_2 \stackrel{def}{=} cond(\mathbf{s}@myloc = 1 \triangleright P_4; true \triangleright P_5)$$
$$P_3 \stackrel{def}{=} \overline{rep}.(p : \sqrt{.P_5} \oplus (1-p) : \overline{rep}.\sqrt{.P_5})$$
$$P_5 \stackrel{def}{=} \sum_{\ell \in \mathbf{Nb}(myloc)} \frac{1}{4} : go\ell.\sqrt{.P}$$

The PRISM encoding of the system follows the translation presented in Section 3. We performed some obvious optimizations in order to reduce the size of our model. All the tests were performed on a G46VW Asus laptop with an Intel i5 2.50 GHz processor and 8 GB of RAM. We ran the tests under Linux Ubuntu 13.04 (Kernel 3.8.0_17), using PRISM 4.0.3 with the MTBDD engine for model checking and CI method for simulation, and Java 7.

As a first experiment we explored and compared the effect of applying policies on the state space of the system in question. Specifically, individuals in the system may engage in two activities: reproduction and dispersal. Let us assume an ordering of these two activities so that reproduction follows dispersal. This gives rise to the policy $\sigma = \{(\tau_{rep,\ell,\mathbf{s}}, \tau_{go,\ell,\mathbf{s}}), (\text{reproduce}_{\ell,\mathbf{s}}, \text{disperse}_{\ell,\mathbf{s}}) \mid \ell \in \mathbf{Loc}\}$.

In Table 5 we summarize the results we obtained in our experiments. In the models we fixed $p = 0.4$. We may observe that applying policy $\sigma$ has resulted in a significant reduction in the size of the state spaces by a factor of 8 on average (see cases *No policy* and *Policy $\sigma$*). A further reduction was achieved by extending our policy in a manner related to the PRISM model: in PRISM each individual is modeled as a discrete module. Thus given an activity of two individuals two distinct executions may arise depending on the order in which the individuals execute the activity. However, we may observe that these two executions both lead to equivalent final states, thus, it is sufficient to consider only one of them. To take this into account, we extended policy $\sigma$ so as to enforce an order on individuals. That is, we require that individuals execute actions in an increasing order in terms of their identifier. This extended policy results in a further reduction of the state space by about 20%.

| Case study size | Number of States | Number of Transitions | Construction time (sec.) | RAM (GB) |
|---|---|---|---|---|
| No policy [37] | | | | |
| 3 PALPS individuals | 130397 | 404734 | 8 | 0.5 |
| 4 PALPS individuals | 1830736 | 7312132 | 101 | 1.9 |
| Policy $\sigma$ | | | | |
| 3 PALPS individuals | 27977 | 64282 | 3 | 0.3 |
| 4 PALPS individuals | 148397 | 409342 | 10 | 0.7 |

| Extended policy | | | | |
|---|---|---|---|---|
| 3 PALPS individuals | 20201 | 41602 | 3 | 0.3 |
| 4 PALPS individuals | 128938 | 310393 | 9 | 0.6 |

**Table 5:** Performance of building probabilistic models in PRISM with and without policies.

As a second experiment, we attempted to determine the limits for *simulating* PALPS models. We constructed PRISM models with various numbers of modules of active and inactive individuals and we run them on PRISM. In Table 6, we summarize the results. It turns out that for models with more than 5000 individuals simulation requires at least 12 hours (which was the time limit we set for our simulations).

| Individuals | File Size (MB) | RAM (GB) | Simulation Time (s) |
|---|---|---|---|
| 10 | 0.1 | 0.18 | 1 |
| 100 | 0.4 | 0.3 | 8 |
| 500 | 2.0 | 0.5 | 45 |
| 1000 | 4.2 | 1.0 | 300 |
| 1500 | 6.2 | 0.7 | 454 |
| 2000 | 8.2 | 0.9 | 820 |
| 5000 | 20.1 | 2.0 | > 12 hours |
| 10000 | 44.1 | 3.4 | > 12 hours |

**Table 6:** Performance of simulating probabilistic systems in PRISM.

Subsequently, we looked into the restriction imposed by our assumption of bounded replication. In particular, this restriction may lead to deadlocks when an active individual attempts to reproduce but no inactive module is available for synchronization. To explore this, we used the *simulation* environment of PRISM and searched for deadlocks by repeating 100 simulations of the model of maximum path length 1000 time steps. Although, this procedure is not complete, we may consider it sufficient as it looks into a fairly large number of life cycles of the population. In Table 7 we summarize the results obtained.

| Active individuals | Inactive individuals | Deadlock ($x,y,z$) |
|---|---|---|
| 3 | 18 | (No,Yes,Yes) |
| 4 | 24 | (No,Yes,Yes) |
| 5 | 30 | (No,Yes,Yes) |
| 6 | 36 | (No,Yes,Yes) |
| 7 | 42 | (No,No,Yes) |
| 8 | 48 | (No,No,Yes) |
| 9 | 56 | (No,No,Yes) |
| 10 | 60 | (No,No,No) |

**Table 7:** Occurrence of deadlock in various instances of the model. The values ($x,y,z$) refer to the presence of deadlock in the case of 4 locations ($x$), 9 locations ($y$), and 16 locations ($z$).

In addition to simulating models in PRISM, we also took advantage of the model checking capabilities of PRISM and, in particular, we checked properties by using the *model-checking by simulation* option, referred to as *confidence interval* CS simulation method (see [1] for more details). We considered several instances of our model consisting of $n$ active individuals, $6 \times n$ inactive individuals and $l$ locations for various values of $n$ and $l$ and we specified to the tool the options of using 100 samples and a confidence interval of 0.01.

The property we experimented with is $R =?[C^{\leq k}]$. This property is a reward-based property that computes the average reward accumulated within the first $k$ execution steps of the model. To check this property, it is necessary to associate rewards with actions of interest within a model. We chose to assign rewards to (1) the clock action and (2) the reproduction actions so as to compute the average number of clock ticks and reproductions that take place within $k$ execution steps of a model. For example, for assigning rewards to the activity of reproduction of the module $P1$, that is, the first active individual, the reward structure is defined as follows:

```
rewards "repP1"
[rep1_n] true : 1;
...
[rep1_n+m] true : 1;
endrewards
```

The value of $k$ for the considered number of execution steps was fixed to $50 \times (n + m)$, where $n + m$ is the total number of individuals (active and inactive) of the model under consideration. In this way, we enabled the model to run on average 50 steps per individual. Figures **??** and **??** summarize the obtained results.

*

(a)

*

(b)

**Fig. 1:** Average number of (a) time steps in PRISM to simulate one time unit in PALPS and (b) reproductions of an individual per time unit.

As another experiment, we computed the percentage of active individuals at the end of the simulation. This is done by verifying the reachability property "eventually, the number of individuals is equal to $s$" for different values of $s$, where $n \leq s \leq m + n$. We performed this analysis for both our model with policy $\sigma$ but also in the variation of our model where the order of the activities of dispersal and reproduction were swapped:

$$\sigma' = \{(\text{dispersal}_{\ell,\mathbf{s}}, \text{reproduction}_{\ell,\mathbf{s}}), (\tau_{go,\ell,\mathbf{s}}, \tau_{rep,\ell,\mathbf{s}}) \mid \ell \in \mathbf{Loc}\}$$

The results are presented in Fig. 2(a) (policy $\sigma$) and Fig. 2(b) (policy $\sigma'$). The results show that the percentages are somewhat higher under the $\sigma'$ policy, especially in the case of the large grid (9 locations) because the bigger the grid the higher the possibility of reproduction and $\sigma'$ allows individuals to reproduce sooner.

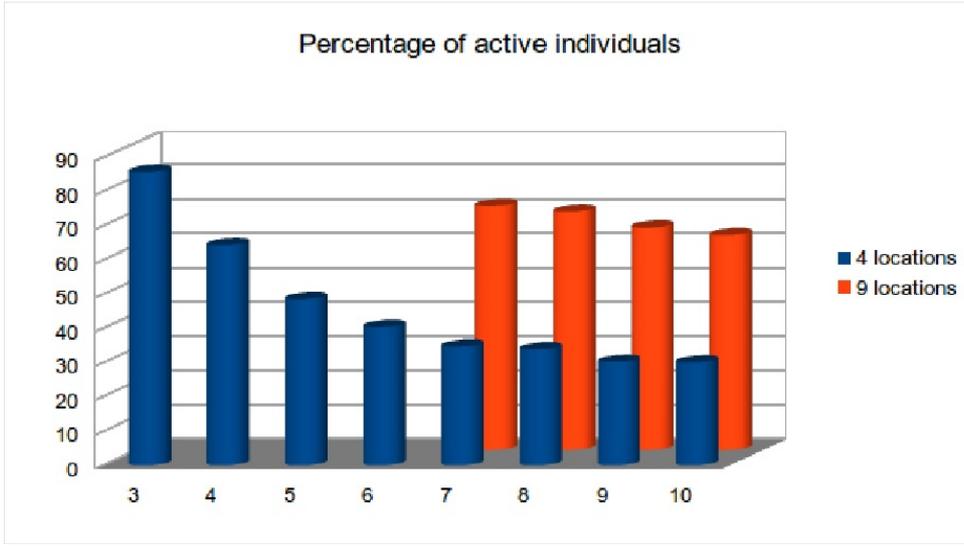

(a)

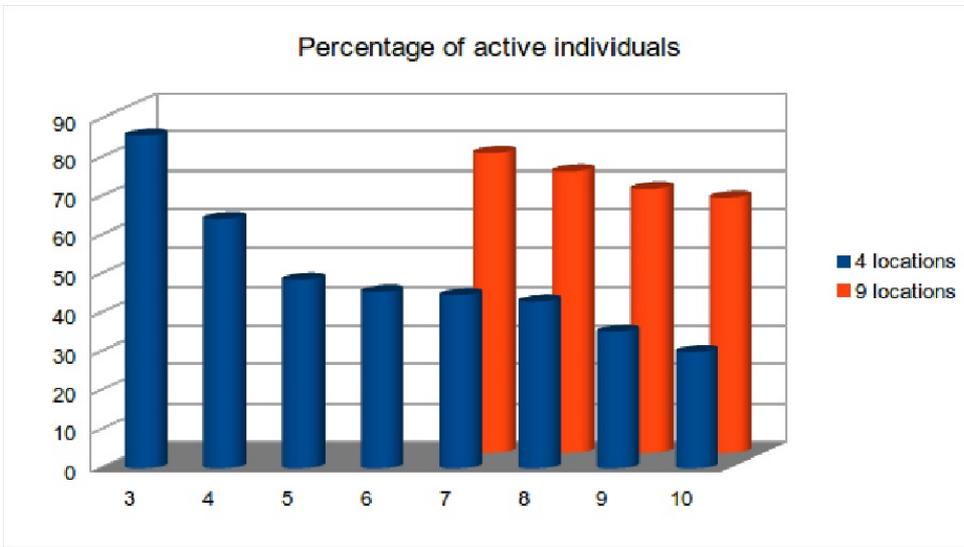

(b)

**Fig. 2:** Percentage of active individuals at the end of the simulation with (a) policy $\sigma$ and (b) policy $\sigma'$.

Consequently, we redeveloped our model of the *Varroa mite* according to the description presented in [42]. In contrast to Example 1, the new model features mortality. Specifically, the new model has two parameters: $b$ the *offspring size* and $p$ the *probability to survive before breeding*. Each mite begins its life by being exposed to death and it survives with a probability $p$. In case of survival, it disperses to a new location. If it has exclusive use of the location then it produces an offspring of size $b$ and it dies. If the location is shared with other mites then all mites die without reproducing. As before, we model space as a lattice with periodic boundary conditions and the probability of dispersal from a location to any of its four neighbors equal to 1/4. As in the previous example, in our system we employed the policy specifying that the process of dispersal precedes reproduction. Formally, the behavior of a mite is defined as follows:

$$P \stackrel{def}{=} p{:}P_1 + (1-p){:}\sqrt{.0}$$
$$P_1 \stackrel{def}{=} \sum_{\ell \in \mathbf{Nb}(myloc)} \frac{1}{4} : go\ell.cond(\mathbf{s}@myloc = 1 \triangleright P_2; true \triangleright \sqrt{.0})$$
$$P_2 \stackrel{def}{=} \overline{rep}^b.\sqrt{.0} \text{ where } \overline{rep}^b \stackrel{def}{=} \underbrace{\overline{rep}...\overline{rep}}_{b \text{ times}}$$

For this model we again checked properties by using the *model-checking by simulation* option. The property we experimented with is $R = ?$ $[I = k]$. This property is a reward-based property that computes the average state instant reward at time $k$. We were interested to study the expected size of the population. For this, we associate to each state a reward representing this size. In our experiments, we varied the size of the initial population ($i$), while the probability of surviving ($p$) and the offspring size ($b$) were fixed to $p = 0.9$ and $b = 3$, and the lattice was of size $4 \times 4$. The number of idle processes was fixed to $n \times b - i$, which is sufficient to avoid deadlocks. The results of the experiments, shown in Fig. 3, demonstrate a tendency of convergence to a stable state and an independence of the initial population for $i > 8$.

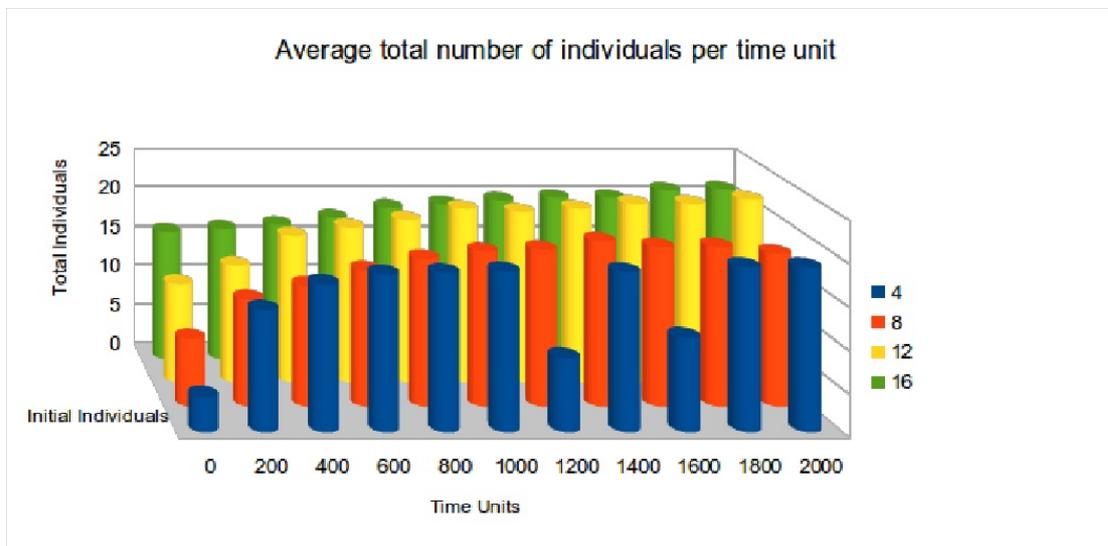

**Fig. 3:** Expected population size vs simulation time for different initial sizes of the population.

We also analyzed, for this model, the effect of the parameters $b$ and $p$ on the evolution of the average total number of individuals through time, with an initial population of 1 individual, as shown in Fig. 4 and Fig. 5. The chosen values for $p$ and $b$ were selected so that they are close to the estimates of the parameters of the Varroa mite, namely, $b = 3$ and $p = 0.9$. Finally, we note that the results may also be applicable to other species that follow the same, so-called scramble-contest behavior such as the bean bruchid that attacks several kinds of beans.

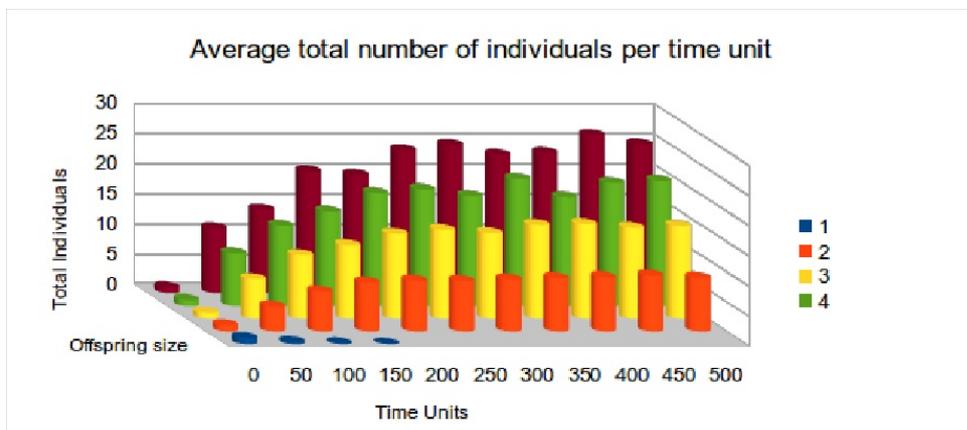

**Fig. 4:** Expected population size vs simulation time for different offspring sizes, for $p = 0.9$ and $i = 1$.

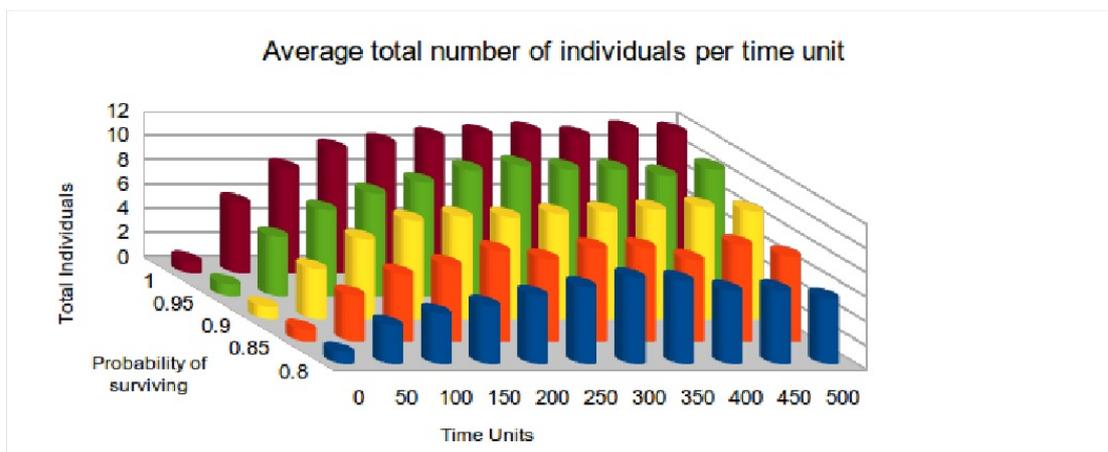

**Fig. 5:** Expected population size vs simulation time for different probabilities of survival, for $b = 3$ and $i = 1$.

## 5 Conclusions

In this paper we have extended the process calculus PALPS with the notion of a *policy*. A policy is an entity that is defined externally to the process-calculus description of an ecological system in order to impose an ordering between the activities taking place within a system as required for the purposes of the analysis. Furthermore, we have described a translation of PALPS *with policies* into the PRISM language. This encoding can be employed for simulating and model checking PALPS systems using the PRISM tool. We experimented with both of these capabilities and we have illustrated types of analysis that can be performed on PALPS models. We have also contrasted our results with those obtained for the same example in our previous work [37]. We have concluded that applying policies can significantly reduce the size of the model thus allowing to consider larger models. For instance, in the example we considered, the state space of the model was reduced by a factor of 10.

As future work, we intend to investigate further approaches for analysis of MDPs that arise from the modeling of population systems. One such approach involves the PRISM tool and concerns the production of PRISM input: we intend to explore alternatives of producing such input possibly via constructing and providing PRISM directly the Markov decision process associated with a PALPS system. We expect that this will result in smaller state spaces than those arising via our PALPS-to-PRISM translation. Furthermore, we would like to explore other directions for reducing the state-space of PALPS models e.g. by enhancing the semantics of PALPS to enable a more succinct presentation of systems especially in terms of the multiplicity of individuals, as well as defining a symbolic semantics which applies a symbolic treatment of environments.

Another direction that we are currently exploring is the application of our methodology to new and complex case studies from the local habitat and the exploration of properties such as extinction (e.g., the expected time until extinction), persistence (e.g., the long-term average of the number of sites occupied at a given time) and spatial indices (e.g., the correlation among nearby locations in space, patch shape analysis and the number of subpopulations in a spatially dispersed metapopulation) similarly to [41].

Another possibility is consider how to apply this techniques to DNA sequencing problems or routing problems similarly to [32, 28, 40].

Finally, an interesting future research direction would be extend the work of [25] towards the development of mean-field analysis to represent the average behavior of systems within a spatially-explicit framework.